\documentclass{article}
\usepackage{amssymb}
\usepackage{amsmath}
\usepackage[dvips]{graphicx}

\input{tcilatex}

\begin{document}

\title{Fermions Tunnelling from Black Holes}
\author{Ryan Kerner\thanks{%
rkerner@uwaterloo.ca} and R.B. Mann\thanks{%
rbmann@sciborg.uwaterloo.ca} \\
Department of Physics \&\ Astronomy, University of Waterloo\\
Waterloo, Ontario N2L 3G1, Canada}
\maketitle

\begin{abstract}
We investigate the tunnelling of spin 1/2 particles through event horizons.
We first apply the tunnelling method to Rindler spacetime and obtain the
Unruh temperature. \ We then apply fermion tunnelling to a general
non-rotating black hole metric and show that the Hawking temperature is
recovered.\ \ 
\end{abstract}

\section{Introduction}

In recent years, a semi-classical method of modeling Hawking radiation as a
tunneling effect has been developed and has garnered a lot of interest \cite%
{early tunnelling}-\cite{Godel}. \ The earliest work with black hole
tunnelling was done by Kraus and Wilczek \cite{early tunnelling}, an
approach that was subsequently refined by various researchers \cite{late 90s
tunnelling,Parikh, Padmanabhan}. From this emerged an alternative way of
understanding black hole radiation. In particular\textbf{\ }one can
calculate the Hawking temperature in a manner independent of traditional
Wick Rotation methods\ or Hawking's original method of modelling
gravitational collapse \cite{hawking}. \ Tunnelling provides not only a
useful verification of thermodynamic properties of black holes but also an
alternate conceptual means for understanding the underlying physical process
of black hole radiation. \ It has been shown to be very robust, having been
successfully applied to a wide variety of exotic spacetimes such as Kerr and
Kerr-Newmann cases \cite{kerr and kerr newman, Zhang and Zhao, first paper},
black rings \cite{Black Rings}, the 3-dimensional BTZ black hole \cite%
{Vanzo, BTZ}, Vaidya \cite{Vaidya}, other dynamical black holes \cite%
{dynamicalbh}, Taub-NUT spacetimes \cite{first paper}, and G\"{o}del
spacetimes \cite{Godel}. \ Tunnelling methods have even been applied to
horizons that are not black hole horizons, such as Rindler Spacetimes \cite%
{Padmanabhan},\cite{first paper} and it has been shown the Unruh temperature 
\cite{Unruh} is in fact recovered.

In general the tunnelling methods\ involve calculating the imaginary part of
the action for the (classically forbidden) process of\ s-wave emission
across the horizon, which in turn is related to the Boltzmann factor for
emission at the Hawking temperature. \ There are two different approaches
that are used to calculate the imaginary part of the action for the emitted
particle. \ The first black hole tunnelling method developed was the Null
Geodesic Method used by Parikh and Wilczek \cite{early tunnelling} which
followed from the work of Kraus and Wilczek \cite{early tunnelling}. \ The
other approach to black hole tunnelling is the Hamilton-Jacobi Ansatz used
by Agheben et al which is an extension of the complex path analysis of
Padmanabhan et al \cite{Padmanabhan}. \ Both of these approaches to
tunnelling use the fact that the WKB approximation of the tunneling
probability for the classically forbidden trajectory from inside to outside
the horizon is given by:%
\begin{equation}
\Gamma \propto \exp (-2\mathrm{Im}I)
\end{equation}%
where $I$ is the classical action of the trajectory to leading order in $%
\hslash $\ (here set equal to unity). \ Where these two methods differ is in
how the action is calculated. For the Null Geodesic method the only part of
the action that contributes an imaginary term is $%
\int_{r_{in}}^{r_{out}}p_{r}dr$, where $p_{r}$is the momentum of the emitted
null s-wave. \ Then by using Hamilton's equation and knowledge of the null
geodesics it is possible to calculate the imaginary part of the action. \
For the Hamilton-Jacobi ansatz it is assumed that the action of the emitted
scalar particle satisfies the relativistic Hamilton-Jacobi equation. From
the symmetries of the metric one picks an appropriate ansatz for the form of
the action and plugs it into the Relativistic Hamilton-Jacobi Equation to
solve. (For a detailed comparison of the Hamilton-Jacobi Ansatz and
Null-Geodesic methods see \cite{first paper}). \ 

Since a black hole has a well defined temperature it should radiate all
types of particles like a black body at that temperature (ignoring grey body
effects). The emission spectrum therefore is expected to contain particles
of all spins; the implications of this expectation were studied 30 years ago 
\cite{don page}. \ However application of tunnelling methods themselves to
date have only involved scalar particles. Specifically there is no other
black hole tunnelling calculation (to the best of our knowledge) that models
fermions tunnelling from the black hole.\ In fact comparatively little has
been done for fermion radiation for black holes. \ The Hawking temperature
for fermion radiation has been calculated for 2d black holes \cite{2d bh
fermion} using the Bogoliubov transformation and more recently was
calculated for evaporating black holes using a technique called the
generalized tortoise coordinate transformation (GTCT) \cite{gtct}-\cite%
{Kinnersley BH}. \ The latter result \cite{Kinnersley BH} is interesting
because there is a contribution to the fermion emission probability due to a
coupling effect between the spin of the emitted fermion and the acceleration
of the Kinnersley black hole. \ From this one may infer that when fermions
are emitted from rotating black holes that will be a coupling between the
spin of the fermion and angular momentum of the rotating black hole present
in the tunnelling probability. \ 

In this paper we extend the tunnelling method to model spin 1/2 particle
emission from non-rotating black holes. \ In order to do this we will follow
an analogous approach to the original approach used by Padmanabhan et al 
\cite{Padmanabhan}. The Hamilton Jacobi ansatz emerged from an application
of the WKB\ approximation to the Klein Gordon equation. \ We will start by
reviewing this general calculation, and then apply a WKB\ approximation to
the Dirac Equation. We consider Rindler spacetime first and confirm that the
Unruh temperature is recovered. Insofar as fermionic vacua are distinct from
bosonic vacua and can lead to distinct physical results \cite{alsingpaper},
this result is non-trivial. We then extend this technique to general 4-D
black hole metric and show the Hawking temperature is recovered. \ We
illustrate this result in several coordinate systems -- Schwarzschild,
Painlev\'{e}, and Kruskal -- to demonstrate that the result is independent
of this choice. \ This last system is particularly interesting since it has
no coordinate singularities at the horizon. That we obtain the expected
Hawking temperature indicates that tunnelling can be understood as a
bona-fide physical phenomenon.

One of the assumptions of our semi-classical calculation is to neglect any
change of angular momentum of the black hole due to the spin of the emitted
particle. \ For zero-angular momentum black holes with mass much larger than
the Planck mass this is a good approximation. Furthermore, statistically
particles of opposite spin will be emitted in equal numbers, yielding no net
change in the angular momentum of the black hole (although second-order
statistical fluctuations will be present). We confirm that spin 1/2 fermions
are also emitted at the Hawking Temperature. This final result, while not
surprising, furnishes an important confirmation of the robustness of the
tunnelling approach. \ 

\section{Review of the Hamilton-Jacobi Ansatz}

We will consider a general (non-extremal) black hole metric of the form: 
\begin{equation}
ds^{2}=-f(r)dt^{2}+\frac{dr^{2}}{g(r)}+C(r)h_{ij}dx^{i}dx^{j}
\label{genericBH}
\end{equation}

The Klein Gordon equation for a scalar field $\phi $ is: 
\begin{equation*}
g^{\mu \nu }\partial _{\mu }\partial _{\nu }\phi -\frac{m^{2}}{\hbar ^{2}}%
\phi =0
\end{equation*}%
Applying the WKB approximation by assuming an ansatz of the form 
\begin{equation*}
\phi (t,r,x^{i})=\exp [\frac{i}{\hbar }I(t,r,x^{i})+I_{1}(t,r,x^{i})+O(\hbar
)]
\end{equation*}%
and then inserting this back into the Klein Gordon equation we get the usual
result of the Hamilton-Jacobi equation to the lowest order in $\hbar $:%
\begin{equation*}
-\left[ g^{\mu \nu }\partial _{\mu }I\partial _{\nu }I+m^{2}\right] +O(\hbar
)=0
\end{equation*}%
(obtained after dividing by the exponential term and multiplying by 
h{\hskip-.2em}\llap{\protect\rule[1.1ex]{.325em}{.1ex}}{\hskip.2em}%
$^{2}$).

For our metric the Hamilton-Jacobi equation is explicitly 
\begin{equation}
-\frac{(\partial _{t}I)^{2}}{f(r)}+g(r)(\partial _{r}I)^{2}+\frac{h^{ij}}{%
C(r)}\partial _{i}I\partial _{j}I+m^{2}=0
\end{equation}%
for the black hole metric (\ref{genericBH}) where we neglect the effects of
the self-gravitation of the particle. There exists a solution of the form 
\begin{equation}
I=-Et+W(r)+J(x^{i})+K
\end{equation}%
where 
\begin{equation*}
\partial _{t}I=-E,\text{ \ \ \ \ \ }\partial _{r}I=W^{\prime }(r),\text{ \ \
\ \ }\partial _{i}I=J_{i}
\end{equation*}%
and $K$ and the $J_{i}$'s are constant ($K$ can be complex). \ Since $%
\partial _{t}$ is the timelike killing vector for this coordinate system, $E$
is the energy.\textbf{\ }Solving for $W(r)$ yields%
\begin{equation}
W_{\pm }(r)=\pm \int \frac{dr}{\sqrt{f(r)g(r)}}\sqrt{E^{2}-f(r)(m^{2}+\frac{%
h^{ij}J_{i}J_{j}}{C(r)})}  \label{wdef}
\end{equation}%
since the equation was quadratic in terms of $W(r)$. \ One solution
corresponds to scalar particles moving away from the black hole (i.e. +
outgoing) and the other solution corresponds to particles moving toward the
black hole (i.e. - incoming). \ \ Imaginary parts of the action can only
come due the pole at the horizon or from the imaginary part of $K$. The
probabilities of crossing the horizon each way are proportional to 
\begin{eqnarray}
\text{Prob}[out] &\propto &\exp [-\frac{2}{\hbar }\func{Im}I]=\exp [-\frac{2%
}{\hbar }(\func{Im}W_{+}+\func{Im}K)]  \label{outprob} \\
\text{Prob}[in] &\propto &\exp [-\frac{2}{\hbar }\func{Im}I]=\exp [-\frac{2}{%
\hbar }(\func{Im}W_{-}+\func{Im}K)]  \label{inprob}
\end{eqnarray}

To ensure that the probability is normalized so that any incoming particles
crossing the horizon have a $100\%$ chance of entering the black hole we set%
\textbf{\ } $\func{Im}K=-\func{Im}W_{-}$ and since $W_{+}=-W_{-}$ this
implies that the probability of a particle tunnelling from inside to outside
the horizon is:%
\begin{equation}
\Gamma \propto \exp [-\frac{4}{\hbar }\func{Im}W_{+}]  \label{finprob}
\end{equation}

Henceforth we set $\hbar $ to unity and also drop the ``$+$'' subscript from 
$W$. Integrating around the pole at the horizon leads to the result \cite%
{first paper} 
\begin{equation}
W=\frac{\pi iE}{\sqrt{g^{\prime }(r_{0})f^{\prime }(r_{0})}}  \label{FormW}
\end{equation}%
where the imaginary part of $W$ is now manifest. This leads to a tunnelling
probability of:%
\begin{equation*}
\Gamma =\exp [-\frac{4\pi }{\sqrt{f^{\prime }(r_{0})g^{\prime }(r_{0})}}E]
\end{equation*}%
and implies the usual Hawking temperature of:%
\begin{equation}
T_{H}=\frac{\sqrt{f^{\prime }(r_{0})g^{\prime }(r_{0})}}{4\pi }
\label{Thvan}
\end{equation}%
It can be shown \cite{Mitra} that the proper Hawking temperature is
recovered for multiple choices of the form of the metric for the same black
hole.

\section{Spin 1/2 particles and Rindler Space}

We first consider the Rindler spacetime, for which the tunnelling
calculation of a scalar field has shown \cite{Padmanabhan},\cite{first paper}
that the Unruh temperature \cite{Unruh} is recovered.

We will only show the calculation explicitly for spin up case; the final
result is also the same for the spin down case as can be easily shown using
the methods described below. \ Due to the statistical nature of the heat
bath we assume that no angular momentum is imparted to the accelerating
detector (i.e. on average there are as many spin up particles as spin down
particles detected). \ The fermionic heat bath as seen by accelerated
observers has many applications, such as understanding the effects of
acceleration on entanglement \cite{Dirac entanglement}.

We will use the following metric for Rindler spacetime 
\begin{eqnarray*}
ds^{2} &=&-f(z)dt^{2}+dx^{2}+dy^{2}+\frac{dz^{2}}{g(z)} \\
f(z) &=&a^{2}z^{2}-1 \\
g(z) &=&\frac{a^{2}z^{2}-1}{a^{2}z^{2}}
\end{eqnarray*}%
\bigskip so chosen for its convenience in extending the technique to normal
black holes. \ The Dirac equation is:%
\begin{equation}
i\gamma ^{\mu }D_{\mu }\psi +\frac{m}{\hbar }\psi =0  \label{Dirac}
\end{equation}%
where: 
\begin{eqnarray*}
D_{\mu } &=&\partial _{\mu }+\Omega _{\mu } \\
\Omega _{\mu } &=&\frac{1}{2}i\Gamma _{\text{ \ }\mu }^{\alpha \text{ \ }%
\beta }\Sigma _{\alpha \beta } \\
\Sigma _{\alpha \beta } &=&\frac{1}{4}i[\gamma ^{\alpha },\gamma ^{\beta }]
\end{eqnarray*}

The $\gamma ^{\mu }$ matrices satisfy $\{\gamma ^{\mu },\gamma ^{\nu
}\}=2g^{\mu \nu }\times 1$. \ There are many different ways to choose the $%
\gamma ^{\mu }$ matrices and we will use the following chiral form:%
\begin{eqnarray*}
\gamma ^{t} &=&\frac{1}{\sqrt{f(z)}}\left( 
\begin{array}{cc}
0 & 1 \\ 
-1 & 0%
\end{array}%
\right) \\
\gamma ^{x} &=&\left( 
\begin{array}{cc}
0 & \sigma ^{1} \\ 
\sigma ^{1} & 0%
\end{array}%
\right) \\
\gamma ^{y} &=&\left( 
\begin{array}{cc}
0 & \sigma ^{2} \\ 
\sigma ^{2} & 0%
\end{array}%
\right) \\
\gamma ^{z} &=&\sqrt{g(z)}\left( 
\begin{array}{cc}
0 & \sigma ^{3} \\ 
\sigma ^{3} & 0%
\end{array}%
\right)
\end{eqnarray*}%
where the $\sigma ^{\prime }s$ are simply the Pauli Sigma Matrices: 
\begin{eqnarray*}
\sigma ^{1} &=&\left( 
\begin{array}{cc}
0 & 1 \\ 
1 & 0%
\end{array}%
\right) \\
\sigma ^{2} &=&\left( 
\begin{array}{cc}
0 & -i \\ 
i & 0%
\end{array}%
\right) \\
\sigma ^{3} &=&\left( 
\begin{array}{cc}
1 & 0 \\ 
0 & -1%
\end{array}%
\right)
\end{eqnarray*}%
and $\xi _{\uparrow /\downarrow }$ are the eigenvectors of $\sigma ^{3}$.
Note that\textbf{\ 
\begin{equation*}
\gamma ^{5}=i\gamma ^{t}\gamma ^{x}\gamma ^{y}\gamma ^{z}=\sqrt{\frac{g(z)}{%
f(z)}}\left( 
\begin{array}{cc}
-1 & 0 \\ 
0 & 1%
\end{array}%
\right)
\end{equation*}%
}is the resulting $\gamma ^{5}$ matrix.

Measuring spin in the z-direction (i.e. the direction of the accelerating
observer)\textbf{\ } we employ the following ansatz for the Dirac field,
respectively corresponding to the spin up and spin down cases: 
\begin{eqnarray}
\psi _{\uparrow }(t,x,y,z) &=&\left[ 
\begin{array}{c}
A(t,x,y,z)\xi _{\uparrow } \\ 
B(t,x,y,z)\xi _{\uparrow }%
\end{array}%
\right] \exp \left[ \frac{i}{\hbar }I_{\uparrow }(t,x,y,z)\right]  \notag \\
&=&\left[ 
\begin{array}{c}
A(t,x,y,z) \\ 
0 \\ 
B(t,x,y,z) \\ 
0%
\end{array}%
\right] \exp \left[ \frac{i}{\hbar }I_{\uparrow }(t,x,y,z)\right]
\label{spin up} \\
\psi _{\downarrow }(t,x,y,z) &=&\left[ 
\begin{array}{c}
C(t,x,y,z)\xi _{\downarrow } \\ 
D(t,x,y,z)\xi _{\downarrow }%
\end{array}%
\right] \exp \left[ \frac{i}{\hbar }I_{\downarrow }(t,x,y,z)\right]  \notag
\\
&=&\left[ 
\begin{array}{c}
0 \\ 
C(t,x,y,z) \\ 
0 \\ 
D(t,x,y,z)%
\end{array}%
\right] \exp \left[ \frac{i}{\hbar }I_{\downarrow }(t,x,y,z)\right]
\label{spin down}
\end{eqnarray}

In order to apply the WKB\ approximation we insert the ansatz \textbf{(}\ref%
{spin up})\ for spin up particles\textbf{\ } into the Dirac Equation.
Dividing by the exponential term and multiplying by $\hbar $ the resulting
equations to leading order in $\hbar $\ are%
\begin{eqnarray}
-B\left( \frac{1}{\sqrt{f(z)}}\partial _{t}I_{\uparrow }+\sqrt{g(z)}\partial
_{z}I_{\uparrow }\right) +Am &=&0  \label{spin1} \\
-B\left( \partial _{x}I_{\uparrow }+i\partial _{y}I_{\uparrow }\right) &=&0
\label{spin2} \\
A\left( \frac{1}{\sqrt{f(z)}}\partial _{t}I_{\uparrow }-\sqrt{g(z)}\partial
_{z}I_{\uparrow }\right) +Bm &=&0  \label{spin3} \\
-A\left( \partial _{x}I_{\uparrow }+i\partial _{y}I_{\uparrow }\right) &=&0
\label{spin4}
\end{eqnarray}%
Note that although $A,B$ are not constant, their derivatives -- and the
components $\Omega _{\mu }$ -- are all of order $O(\hbar )$ and so can be
neglected to lowest order in WKB.

When $m\neq 0$ equations (\ref{spin1}) and (\ref{spin3}) couple whereas when 
$m=0$ they decouple. We employ the ansatz 
\begin{equation}
I_{\uparrow }=-Et+W(z)+P(x,y)  \label{spinansatz}
\end{equation}%
and insert it into equations (\ref{spin1}-\ref{spin4}) 
\begin{eqnarray}
-B\left( \frac{-E}{\sqrt{f(z)}}+\sqrt{g(z)}W^{\prime }(z)\right) +mA &=&0
\label{spin5} \\
-B\left( P_{x}+iP_{y}\right) &=&0  \label{spin6} \\
-A\left( \frac{E}{\sqrt{f(z)}}+\sqrt{g(z)}W^{\prime }(z)\right) +mB &=&0
\label{spin7} \\
-A\left( P_{x}+iP_{y}\right) &=&0  \label{spin8}
\end{eqnarray}%
where we consider only the positive frequency contributions without loss of%
\textbf{\ }generality. Equations (\ref{spin6}) and (\ref{spin8}) both yield $%
\left( P_{x}+iP_{y}\right) =0$ regardless of $A$ or $B$, implying 
\begin{equation}
P(x,y)=h(x+iy)  \label{Psoln}
\end{equation}%
where $h$ is some arbitrary function.

Consider first $m=0$. Equations (\ref{spin5}) and (\ref{spin7}) then have
two possible solutions\textbf{\ }%
\begin{eqnarray*}
A &=&0\text{ and }W^{\prime }(z)=W_{+}^{\prime }(z)=\frac{E}{\sqrt{f(z)g(z)}}
\\
B &=&0\text{ and }W^{\prime }(z)=W_{-}^{\prime }(z)=\frac{-E}{\sqrt{f(z)g(z)}%
}
\end{eqnarray*}%
corresponding to motion away from (+) and toward (-) the horizon, chosen to
be at $z=1/a$.

Since the solution $[A,0,0,0]$ \ is an eigenvector of $\gamma ^{5}$ and has
a negative eigenvalue its spin and momentum vectors are opposite, which is
consistent with the fact that the particle is moving toward the horizon and
the spin is up. The solution $[0,0,B,0]$\ is also an eigenvector of $\gamma
^{5}$ with positive eigenvalue; its spin and momentum vectors are therefore
in the same direction, consistent with the particle being spin up and moving
away from the horizon.\textbf{\ }

Hence with the Rindler horizon at $z=1/a$ the ($\pm $) cases correspond to
outgoing/incoming solutions of the same spin. Note that neither of these
cases is an antiparticle solution since we assumed positive frequency modes
as a part of the ansatz. In computing the imaginary part of the action we
note that $P(x,y)$\ must be complex (other than the trivial solution of $P=0$%
), and so will yield a contribution. However it is the same for both
incoming and outgoing solutions, and so will cancel out in computing the
emission probability 
\begin{eqnarray}
\Gamma &\propto &\frac{\text{Prob}[out]}{\text{Prob}[in]}=\frac{\exp [-2(%
\func{Im}W_{+}+\func{Im}h)]}{\exp [-2(\func{Im}W_{-}+\func{Im}h)]} \\
&=&\exp [-2(\func{Im}W_{+}-\func{Im}W_{-})=\exp [-4\func{Im}W_{+}]
\end{eqnarray}%
using reasoning similar to the scalar case. We obtain 
\begin{equation*}
W_{+}(z)=\int \frac{Edz}{\sqrt{f(z)g(z)}}
\end{equation*}%
and after integrating around the pole (and dropping the + subscript) 
\begin{equation}
W=\frac{\pi iE}{\sqrt{g^{\prime }(z_{0})f^{\prime }(z_{0})}}=\frac{\pi iE}{2a%
}
\end{equation}%
The resulting tunnelling probability is 
\begin{equation*}
\Gamma =\exp [-\frac{2\pi }{a}E]
\end{equation*}%
recovering 
\begin{equation}
T_{H}=\frac{a}{2\pi }  \label{unruhtemp}
\end{equation}%
which is the Unruh temperature.

In the massive case equations (\ref{spin5}) and (\ref{spin7}) no longer
decouple. \ We will start by eliminating the function $W^{\prime }(z)$ from
the two equations so we can find an equation relating $A$ and $B$ in terms
of known values. Subtracting $B\times $(\ref{spin7}) from $A\times $\ (\ref%
{spin5}) gives 
\begin{eqnarray*}
\frac{2ABE}{\sqrt{f(z)}}+mA^{2}-mB^{2} &=&0 \\
m\sqrt{f(z)}(\frac{A}{B})^{2}+2E(\frac{A}{B})-m\sqrt{f(z)} &=&0
\end{eqnarray*}%
and so 
\begin{equation*}
\frac{A}{B}=\frac{-E\pm \sqrt{E^{2}+m^{2}f(z)}}{m\sqrt{f(z)}}
\end{equation*}%
where 
\begin{eqnarray*}
\lim_{z\rightarrow z_{0}}\left( \frac{-E+\sqrt{E^{2}+m^{2}f(z)}}{m\sqrt{f(z)}%
}\right) &=&0 \\
\lim_{z\rightarrow z_{0}}\left( \frac{-E-\sqrt{E^{2}+m^{2}f(z)}}{m\sqrt{f(z)}%
}\right) &=&-\infty
\end{eqnarray*}

Consequently at the Rindler horizon either $\frac{A}{B}\rightarrow 0$ or $%
\frac{A}{B}\rightarrow -\infty $, i.e. either $A\rightarrow 0$ or $%
B\rightarrow 0$. For $A\rightarrow 0$ at the horizon, we solve (\ref{spin7})
in terms of $m$ and insert into (\ref{spin5}) 
\begin{eqnarray*}
-B\left( \frac{-E}{\sqrt{f(z)}}+\sqrt{g(z)}W^{\prime }(z)\right) +\frac{A^{2}%
}{B}\left( \frac{E}{\sqrt{f(z)}}+\sqrt{g(z)}W^{\prime }(z)\right) &=&0 \\
\frac{EB}{\sqrt{f(z)}}\left( 1+\frac{A^{2}}{B^{2}}\right) -B\sqrt{g(z)}%
W^{\prime }(z)\left( 1-\frac{A^{2}}{B^{2}}\right) &=&0
\end{eqnarray*}%
\begin{equation*}
W^{\prime }(z)=W_{+}^{\prime }(z)=\frac{E}{\sqrt{f(z)g(z)}}\frac{\left( 1+%
\frac{A^{2}}{B^{2}}\right) }{\left( 1-\frac{A^{2}}{B^{2}}\right) }
\end{equation*}%
whereas for $B\rightarrow 0$ at the horizon we solve (\ref{spin5}) in terms
of $m$ and insert into (\ref{spin7}) to get 
\begin{eqnarray*}
-A\left( \frac{E}{\sqrt{f(z)}}+\sqrt{g(z)}W^{\prime }(z)\right) +\frac{B^{2}%
}{A}\left( \frac{-E}{\sqrt{f(z)}}+\sqrt{g(z)}W^{\prime }(z)\right) &=&0 \\
-\frac{EA}{\sqrt{f(z)}}\left( 1+\frac{B^{2}}{A^{2}}\right) -A\sqrt{g(z)}%
W^{\prime }(z)\left( 1-\frac{B^{2}}{A^{2}}\right) &=&0
\end{eqnarray*}

\begin{equation*}
W^{\prime }(z)=W_{-}^{\prime }(z)=\frac{-E}{\sqrt{f(z)g(z)}}\frac{\left( 1+%
\frac{B^{2}}{A^{2}}\right) }{\left( 1-\frac{B^{2}}{A^{2}}\right) }
\end{equation*}

Since the extra contributions vanish at the horizon, the result of
integrating around the pole for $W$ in the massive case is the same as the
massless case and we recover the Unruh temperature for the fermionic Rindler
vacuum.

The spin-down case proceeds in a manner fully analogous to the spin-up case
discussed above. Other than some changes of sign the equations are of the
same form as the spin up case. For both the massive and massless cases the
Unruh temperature (\ref{unruhtemp}) is obtained, implying that both spin up
and spin down particles are emitted at the same rate.

\section{Black Hole Fermion Emission}

We turn next to a general static spherically symmetric black hole. As stated
in the introduction, we will ignore any change in the angular momentum of
the black hole due to the spin of the emitted particle. \ This is a good
approximation for black holes of sufficient mass. The zero angular momentum
state is maintained because statistically as many particles with spin in one
direction will be emitted as particles with spin in the opposite direction.

We will now extend the fermion tunnelling approach to a general black hole
with spherical symmetry. The metric is 
\begin{equation}
ds^{2}=-f(r)dt^{2}+\frac{dr^{2}}{g(r)}+r^{2}(d\theta ^{2}+\sin ^{2}(\theta
)d\phi ^{2})  \label{sssmet}
\end{equation}%
where for this case we will pick for the $\gamma $ matrices 
\begin{eqnarray*}
\gamma ^{t} &=&\frac{1}{\sqrt{f(r)}}\left( 
\begin{array}{cc}
i & 0 \\ 
0 & -i%
\end{array}%
\right) \\
\gamma ^{r} &=&\sqrt{g(r)}\left( 
\begin{array}{cc}
0 & \sigma ^{3} \\ 
\sigma ^{3} & 0%
\end{array}%
\right) \\
\gamma ^{\theta } &=&\frac{1}{r}\left( 
\begin{array}{cc}
0 & \sigma ^{1} \\ 
\sigma ^{1} & 0%
\end{array}%
\right) \\
\gamma ^{\phi } &=&\frac{1}{r\sin \theta }\left( 
\begin{array}{cc}
0 & \sigma ^{2} \\ 
\sigma ^{2} & 0%
\end{array}%
\right)
\end{eqnarray*}%
where we measure spin in terms of the $r$-direction. The matrix for $\gamma
^{5}$ is 
\begin{equation*}
\gamma ^{5}=i\gamma ^{t}\gamma ^{r}\gamma ^{\theta }\gamma ^{\phi }=i\sqrt{%
\frac{g(r)}{f(r)}}\frac{1}{r^{2}\sin \theta }\left( 
\begin{array}{cc}
0 & -1 \\ 
1 & 0%
\end{array}%
\right)
\end{equation*}

The spin up (i.e. +ve $r$-direction) and spin down (i.e. -ve $r$-direction)
solutions have the form 
\begin{eqnarray}
\psi _{\uparrow }(t,r,\theta ,\phi ) &=&\left[ 
\begin{array}{c}
A(t,r,\theta ,\phi )\xi _{\uparrow } \\ 
B(t,r,\theta ,\phi )\xi _{\uparrow }%
\end{array}%
\right] \exp \left[ \frac{i}{\hbar }I_{\uparrow }(t,r,\theta ,\phi )\right] 
\notag \\
&=&\left[ 
\begin{array}{c}
A(t,r,\theta ,\phi ) \\ 
0 \\ 
B(t,r,\theta ,\phi ) \\ 
0%
\end{array}%
\right] \exp \left[ \frac{i}{\hbar }I_{\uparrow }(t,r,\theta ,\phi )\right]
\label{spinupbh} \\
\psi _{\downarrow }(t,x,y,z) &=&\left[ 
\begin{array}{c}
C(t,r,\theta ,\phi )\xi _{\downarrow } \\ 
D(t,r,\theta ,\phi )\xi _{\downarrow }%
\end{array}%
\right] \exp \left[ \frac{i}{\hbar }I_{\downarrow }(t,r,\theta ,\phi )\right]
\notag \\
&=&\left[ 
\begin{array}{c}
0 \\ 
C(t,r,\theta ,\phi ) \\ 
0 \\ 
D(t,r,\theta ,\phi )%
\end{array}%
\right] \exp \left[ \frac{i}{\hbar }I_{\downarrow }(t,r,\theta ,\phi )\right]
\label{spindnbh}
\end{eqnarray}

We will only solve the spin up case explicitly since the spin-down case is
fully analogous. Employing the ansatz (\ref{spinupbh}) into the Dirac
equation results in 
\begin{eqnarray}
-\left( \frac{iA}{\sqrt{f(r)}}\partial _{t}I_{\uparrow }+B\sqrt{g(r)}%
\partial _{r}I_{\uparrow }\right) +Am &=&0 \\
-\frac{B}{r}\left( \partial _{\theta }I_{\uparrow }+\frac{1}{\sin \theta }%
i\partial _{\phi }I_{\uparrow }\right) &=&0 \\
\left( \frac{iB}{\sqrt{f(r)}}\partial _{t}I_{\uparrow }-A\sqrt{g(r)}\partial
_{r}I_{\uparrow }\right) +Bm &=&0 \\
-\frac{A}{r}\left( \partial _{\theta }I_{\uparrow }+\frac{1}{\sin \theta }%
i\partial _{\phi }I_{\uparrow }\right) &=&0
\end{eqnarray}%
\newline
to leading order in $\hbar $. We assume the action takes the form\textbf{\ }%
\begin{equation}
I_{\uparrow }=-Et+W(r)+J(\theta ,\phi )  \label{actbh}
\end{equation}%
where we only concern ourselves with positive frequency contributions as
before. This yields 
\begin{eqnarray}
\left( \frac{iAE}{\sqrt{f(r)}}-B\sqrt{g(r)}W^{\prime }(r)\right) +mA &=&0
\label{bhspin5} \\
-\frac{B}{r}\left( J_{\theta }+\frac{1}{\sin \theta }iJ_{\phi }\right) &=&0
\label{bhspin6} \\
-\left( \frac{iBE}{\sqrt{f(r)}}+A\sqrt{g(r)}W^{\prime }(r)\right) +Bm &=&0
\label{bhspin7} \\
-\frac{A}{r}\left( J_{\theta }+\frac{1}{\sin \theta }iJ_{\phi }\right) &=&0
\label{bhspin8}
\end{eqnarray}

Notice that (\ref{bhspin6}) and (\ref{bhspin8}) result in the same equation
regardless of $A$ or $B$ (i.e. $\left( J_{\theta }+\frac{1}{\sin \theta }%
iJ_{\phi }\right) =0$ must be true), implying that $J(\theta ,\phi )$ must
be a complex function.\ \ As with the Rindler case, the same solution for $J$%
\ is obtained for both the outgoing and incoming cases. \ Consequently the
contribution from $J$\ cancels out upon dividing the outgoing probability by
the incoming probability as in eq. (\ref{inprob}). \ We therefore can ignore 
$J$\ from this point (or else pick the trivial $J=0$\ solution).

Equations (\ref{bhspin5}) and (\ref{bhspin7}) (for $m=0$) have two possible
solutions: 
\begin{eqnarray*}
A &=&-iB\text{ and }W^{\prime }(r)=W_{+}^{\prime }(r)=\frac{E}{\sqrt{f(r)g(r)%
}} \\
A &=&iB\text{ and }W^{\prime }(r)=W_{-}^{\prime }(r)=\frac{-E}{\sqrt{f(r)g(r)%
}}
\end{eqnarray*}%
where $W_{+}$ corresponds to outward solutions and $W_{-}$ correspond to the
incoming solutions. The overall tunnelling probability is 
\begin{equation}
\Gamma =\frac{\text{Prob}[out]}{\text{Prob}[in]}=\frac{\exp [-2(\func{Im}%
W_{+})]}{\exp [-2(\func{Im}W_{-})]}=\exp [-4\func{Im}W_{+}]
\label{finfermprob}
\end{equation}%
with 
\begin{equation*}
W_{+}(r)=\int \frac{Edr}{\sqrt{f(r)g(r)}}
\end{equation*}%
After integrating around the pole (and dropping the + subscript) we find 
\begin{equation}
W=\frac{\pi iE}{\sqrt{g^{\prime }(r_{0})f^{\prime }(r_{0})}}
\end{equation}%
giving 
\begin{equation}
\Gamma =\exp [-\frac{4\pi }{\sqrt{g^{\prime }(r_{0})f^{\prime }(r_{0})}}E]
\label{Gamfermprob}
\end{equation}%
for the resultant tunnelling probability to leading order in $\hbar $.

We therefore recover the expected Hawking Temperature 
\begin{equation}
T_{H}=\frac{\sqrt{f^{\prime }(r_{0})g^{\prime }(r_{0})}}{4\pi }
\label{Hawking T}
\end{equation}%
in the massless case.

Solving equations (\ref{bhspin5}) and (\ref{bhspin7}) for $A$ and $B$ in the
case that $m\neq 0$ leads to the result:%
\begin{equation*}
\left( \frac{A}{B}\right) ^{2}=\frac{-iE+\sqrt{f(r)}m}{iE+\sqrt{f(r)}m}
\end{equation*}%
and approaching the horizon we see that $\lim_{r\rightarrow r_{0}}\left( 
\frac{A}{B}\right) ^{2}=-1$. Following a procedure similar to what was done
above, we obtain the same result for the Hawking Temperature as in the
massless case.

The spin-down calculation is very similar to the spin-up case discussed
above. Other than some changes of sign, the equations are of the same form
as the spin up case. For both the massive and massless spin down cases the
Hawking temperature (\ref{Hawking T}) is obtained, implying that both spin
up and spin down particles are emitted at the same rate. \ This is
consistent with our initial assumption that there are as many spin up as
spin down fermions emitted. \ 

\subsection{Painlev\'{e} Coordinates}

In this section we demonstrate that Painlev\'{e} coordinates can be used to
recover the results of the preceding section, albeit by a somewhat different
computational route.

Using the transformation 
\begin{equation}
t\rightarrow t-\int \sqrt{\frac{1-g\left( r\right) }{f\left( r\right)
g\left( r\right) }}dr  \label{paintrans}
\end{equation}%
we obtain from the metric (\ref{sssmet}) 
\begin{equation}
ds^{2}=-f(r)dt^{2}+2\sqrt{f(r)}\sqrt{\frac{1}{g(r)}-1}drdt+dr^{2}+r^{2}d%
\Omega ^{2}  \label{painform}
\end{equation}%
which is the Painlev\'{e} form of a spherically symmetric metric.

This coordinate system has a number of interesting features. At any fixed
time the spatial geometry is flat. \ At any fixed radius the boundary
geometry for the Painlev\'{e} metric is exactly the same as that of the
unaltered black hole metric. \ Also, this form of the Painlev\'{e} metric is
a very convenient metric to use for black hole tunnelling since the
imaginary part of the action for the incoming solution is zero which means
Prob$[in]=1$ \cite{Mitra}. \ This property also holds for fermion
tunnelling. \ 

We choose the representation for the $\gamma $ matrices to be 
\begin{eqnarray*}
\gamma ^{t} &=&\frac{1}{\sqrt{f(r)}}\left( 
\begin{array}{cc}
0 & 1+\sqrt{1-g(r)}\sigma ^{3} \\ 
-1+\sqrt{1-g(r)}\sigma ^{3} & 0%
\end{array}%
\right) \\
\gamma ^{r} &=&\sqrt{g(r)}\left( 
\begin{array}{cc}
0 & \sigma ^{3} \\ 
\sigma ^{3} & 0%
\end{array}%
\right) \\
\gamma ^{\theta } &=&\frac{1}{r}\left( 
\begin{array}{cc}
0 & \sigma ^{1} \\ 
\sigma ^{1} & 0%
\end{array}%
\right) \\
\gamma ^{\phi } &=&\frac{1}{r\sin \theta }\left( 
\begin{array}{cc}
0 & \sigma ^{2} \\ 
\sigma ^{2} & 0%
\end{array}%
\right)
\end{eqnarray*}

\ 

The matrix for $\gamma ^{5}$ for this case is:%
\begin{equation*}
\gamma ^{5}=i\gamma ^{t}\gamma ^{r}\gamma ^{\theta }\gamma ^{\phi }=\sqrt{%
\frac{g(r)}{f(r)}}\frac{1}{r^{2}\sin \theta }\left( 
\begin{array}{cc}
-1-\sqrt{1-g(r)}\sigma ^{3} & 0 \\ 
0 & +1-\sqrt{1-g(r)}\sigma ^{3}%
\end{array}%
\right)
\end{equation*}%
Measuring spin in the $r$-direction we have as before the two following
ansatz for the spin 1/2 Dirac field which correspond to the spin up (i.e.
+ve $r$-direction) and spin down (i.e. -ve $r$-direction) cases: 
\begin{eqnarray}
\psi _{\uparrow }(t,r,\theta ,\phi ) &=&\left[ 
\begin{array}{c}
A(t,r,\theta ,\phi )\xi _{\uparrow } \\ 
B(t,r,\theta ,\phi )\xi _{\uparrow }%
\end{array}%
\right] \exp \left[ \frac{i}{\hbar }I_{\uparrow }(t,r,\theta ,\phi )\right] 
\notag \\
&=&\left[ 
\begin{array}{c}
A(t,r,\theta ,\phi ) \\ 
0 \\ 
B(t,r,\theta ,\phi ) \\ 
0%
\end{array}%
\right] \exp \left[ \frac{i}{\hbar }I_{\uparrow }(t,r,\theta ,\phi )\right]
\\
\psi _{\downarrow }(t,x,y,z) &=&\left[ 
\begin{array}{c}
C(t,r,\theta ,\phi )\xi _{\downarrow } \\ 
D(t,r,\theta ,\phi )\xi _{\downarrow }%
\end{array}%
\right] \exp \left[ \frac{i}{\hbar }I_{\downarrow }(t,r,\theta ,\phi )\right]
\notag \\
&=&\left[ 
\begin{array}{c}
0 \\ 
C(t,r,\theta ,\phi ) \\ 
0 \\ 
D(t,r,\theta ,\phi )%
\end{array}%
\right] \exp \left[ \frac{i}{\hbar }I_{\downarrow }(t,r,\theta ,\phi )\right]
\end{eqnarray}

Once again we will only solve the spin up case explicitly. Insertion of the
ansatz into the Dirac equation results in the following equations to the
leading order in $\hbar $. 
\begin{eqnarray}
-B\left( \frac{1}{\sqrt{f(r)}}\left( 1+\sqrt{1-g(r)}\right) \partial
_{t}I_{\uparrow }+\sqrt{g(r)}\partial _{r}I_{\uparrow }\right) +Am &=&0 \\
-\frac{B}{r}\left( \partial _{\theta }I_{\uparrow }+\frac{1}{\sin \theta }%
i\partial _{\phi }I_{\uparrow }\right) &=&0 \\
A\left( \frac{1}{\sqrt{f(r)}}\left( 1-\sqrt{1-g(r)}\right) \partial
_{t}I_{\uparrow }-\sqrt{g(r)}\partial _{r}I_{\uparrow }\right) +Bm &=&0 \\
-\frac{A}{r}\left( \partial _{\theta }I_{\uparrow }+\frac{1}{\sin \theta }%
i\partial _{\phi }I_{\uparrow }\right) &=&0
\end{eqnarray}%
\newline
To solve these equations we pick the ansatz (\ref{actbh}) for the action,
again working only with positive frequency contributions. The equations for $%
J$ are the same as in the last section, and we can dispense with this
function for the same reasons as before. We obtain 
\begin{eqnarray}
B\left( \frac{1}{\sqrt{f(r)}}\left( 1+\sqrt{1-g(r)}\right) E-\sqrt{g(r)}%
W^{\prime }(r)\right) +Am &=&0  \label{bhpspin1} \\
-A\left( \frac{1}{\sqrt{f(r)}}\left( 1-\sqrt{1-g(r)}\right) E+\sqrt{g(r)}%
W^{\prime }(r)\right) +Bm &=&0  \label{bhpspin2}
\end{eqnarray}%
\ \bigskip

Equations (\ref{bhpspin1}) and (\ref{bhpspin2}) (for $m=0$) have two
possible solutions: 
\begin{eqnarray*}
A &=&0\text{ and }W^{\prime }(r)=W_{+}^{\prime }(r)=\frac{E\left( 1+\sqrt{%
1-g(r)}\right) }{\sqrt{f(r)g(r)}} \\
B &=&0\text{ and }W^{\prime }(r)=W_{-}^{\prime }(r)=\frac{-E\left( 1-\sqrt{%
1-g(r)}\right) }{\sqrt{f(r)g(r)}}
\end{eqnarray*}%
$W_{+}$ corresponds to outward solutions and $W_{-}$ correspond to the
incoming solutions. \ Notice that $W_{+}^{\prime }$ have a pole at the
horizon but $W_{-}^{\prime }$ has a well defined limit at the horizon and
does not have a pole (i.e. $\lim_{r\rightarrow r_{0}}W_{-}^{\prime }(r)=%
\frac{-E}{2}\sqrt{\frac{g^{\prime }(r_{0})}{f^{\prime }(r_{0})}}$). \ This
implies that the the imaginary part $W_{-}$ is zero and confirms that Prob$%
[in]=1$. \ So the overall tunnelling probability is: 
\begin{eqnarray*}
\Gamma &\propto &\text{Prob}[out] \\
\Gamma &\propto &\exp [-2\func{Im}W_{+}]
\end{eqnarray*}

$\therefore $

\begin{equation*}
W_{+}(r)=\int \frac{E\left( 1+\sqrt{1-g(r)}\right) dr}{\sqrt{f(r)g(r)}}
\end{equation*}%
and after integrating around the pole (and dropping the + subscript): 
\begin{equation}
W=\frac{2\pi iE}{\sqrt{g^{\prime }(r_{0})f^{\prime }(r_{0})}}
\end{equation}

So the resulting tunnelling probability is once again:%
\begin{equation*}
\Gamma =\exp [-\frac{4\pi }{\sqrt{g^{\prime }(r_{0})f^{\prime }(r_{0})}}E]
\end{equation*}%
and the normal Hawking Temperature is also recovered for the Painlev\'{e}
massless case%
\begin{equation}
T_{H}=\frac{\sqrt{f^{\prime }(r_{0})g^{\prime }(r_{0})}}{4\pi }
\end{equation}

Solving equations (\ref{bhpspin1}) and (\ref{bhpspin2}) for $A$ and $B$ in
the case that $m\neq 0$ leads to the results that $A\rightarrow 0$ as $%
r\rightarrow r_{0}$ or $B\rightarrow 0$ as $r\rightarrow r_{0}$. \ So the
same final result will be recovered in the massive case.

\subsection{Kruskal-Szekers Metric}

In the preceding subsections we employed metrics that had co-ordinate
singularities at the horizon. \ One might be concerned that the tunnelling
effect is dependent upon this. \ Here we demonstrate that this is not the
case, by investigating fermion tunnelling in the Kruskal-Szekers metric%
\begin{equation}
ds^{2}=f(r)\left( -dT^{2}+dX^{2}\right) +r^{2}d\Omega ^{2}  \label{ksmet}
\end{equation}%
where:%
\begin{equation*}
f(r)=\frac{32M^{3}e^{-\frac{r}{2M}}}{r}\text{ \ \ \ \ \ \ \ \ \ \ \ \ }(%
\frac{r}{2M}-1)e^{r/2M}=X^{2}-T^{2}
\end{equation*}%
The metric (\ref{ksmet}) is well behaved at both the future and past
horizons $X=\pm T$ (corresponding to $r=2M$). \ Note that the metric has a
timelike Killing vector $X\partial _{T}+T\partial _{X}$ (and not $\partial
_{T}$). \ 

For this calculation we will employ the following representation for the $%
\gamma $ matrices 
\begin{eqnarray*}
\gamma ^{T} &=&\frac{1}{\sqrt{f(r)}}\left( 
\begin{array}{cc}
0 & 1 \\ 
-1 & 0%
\end{array}%
\right) \\
\gamma ^{X} &=&\frac{1}{\sqrt{f(r)}}\left( 
\begin{array}{cc}
0 & \sigma ^{3} \\ 
\sigma ^{3} & 0%
\end{array}%
\right) \\
\gamma ^{\theta } &=&\frac{1}{r}\left( 
\begin{array}{cc}
0 & \sigma ^{1} \\ 
\sigma ^{1} & 0%
\end{array}%
\right) \\
\gamma ^{\phi } &=&\frac{1}{r\sin \theta }\left( 
\begin{array}{cc}
0 & \sigma ^{2} \\ 
\sigma ^{2} & 0%
\end{array}%
\right)
\end{eqnarray*}%
where we measure spin referenced to the $X$-direction. The matrix for $%
\gamma ^{5}$ is 
\begin{equation*}
\gamma ^{5}=i\gamma ^{t}\gamma ^{r}\gamma ^{\theta }\gamma ^{\phi }=\frac{1}{%
f(r)}\frac{1}{r^{2}\sin \theta }\left( 
\begin{array}{cc}
-1 & 0 \\ 
0 & 1%
\end{array}%
\right)
\end{equation*}

The spin up (i.e. +ve $X$-direction) and spin down (i.e. -ve $X$-direction)
solutions have the form 
\begin{eqnarray}
\psi _{\uparrow }(T,X,\theta ,\phi ) &=&\left[ 
\begin{array}{c}
A(T,X,\theta ,\phi )\xi _{\uparrow } \\ 
B(T,X,\theta ,\phi )\xi _{\uparrow }%
\end{array}%
\right] \exp \left[ \frac{i}{\hbar }I_{\uparrow }(T,X,\theta ,\phi )\right] 
\notag \\
&=&\left[ 
\begin{array}{c}
A(T,X,\theta ,\phi ) \\ 
0 \\ 
B(T,X,\theta ,\phi ) \\ 
0%
\end{array}%
\right] \exp \left[ \frac{i}{\hbar }I_{\uparrow }(T,X,\theta ,\phi )\right]
\label{spinupbhk} \\
\psi _{\downarrow }(T,X,y,z) &=&\left[ 
\begin{array}{c}
C(T,X,\theta ,\phi )\xi _{\downarrow } \\ 
D(T,X,\theta ,\phi )\xi _{\downarrow }%
\end{array}%
\right] \exp \left[ \frac{i}{\hbar }I_{\downarrow }(T,X,\theta ,\phi )\right]
\notag \\
&=&\left[ 
\begin{array}{c}
0 \\ 
C(T,X,\theta ,\phi ) \\ 
0 \\ 
D(T,X,\theta ,\phi )%
\end{array}%
\right] \exp \left[ \frac{i}{\hbar }I_{\downarrow }(T,X,\theta ,\phi )\right]
\label{spindnbhk}
\end{eqnarray}%
Once again inserting the spin-up ansatz (\ref{spinupbhk}) (the spin-down
case being similar) into the Dirac equation yields the following equations 
\begin{eqnarray}
-\frac{B}{\sqrt{f(r)}}\left( \partial _{T}I_{\uparrow }+\partial
_{X}I_{\uparrow }\right) +Am &=&0  \label{kruskal1} \\
-\frac{B}{r}\left( \partial _{\theta }I_{\uparrow }+\frac{1}{\sin \theta }%
i\partial _{\phi }I_{\uparrow }\right) &=&0 \\
\frac{A}{\sqrt{f(r)}}\left( \partial _{T}I_{\uparrow }-\partial
_{X}I_{\uparrow }\right) +Bm &=&0  \label{kruskal2} \\
-\frac{A}{r}\left( \partial _{\theta }I_{\uparrow }+\frac{1}{\sin \theta }%
i\partial _{\phi }I_{\uparrow }\right) &=&0
\end{eqnarray}%
\newline
to leading order in $\hbar $. This time we can infer only that the action
takes the form\textbf{\ }%
\begin{equation}
I_{\uparrow }=-I(X,T)+J(\theta ,\phi )  \label{newaction}
\end{equation}%
The equations for $J$ are unchanged from previous calculations. We thus
ignore these equations since they do not affect the final result and only
concern ourselves with solving for $I(X,T)$. \ 

In order to solve the equations we need a definition of the energy of the
wave. \ We will define energy via the timelike killing vector 
\begin{equation*}
\partial _{\chi }=N(X\partial _{T}+T\partial _{X})
\end{equation*}%
where $N$ is a normalization constant chosen so that the norm of the Killing
vector is equal to $1$ at infinity. \ This yields

\begin{equation}
\partial _{\chi }=\frac{1}{4M}(X\partial _{T}+T\partial _{X})  \label{kskv}
\end{equation}%
and so

\begin{equation}
\partial _{\chi }I=-E  \label{defenergy}
\end{equation}%
Using (\ref{defenergy}) with (\ref{kruskal1}) and (\ref{kruskal2}) we shall
solve the equations.

Consider first the massless case. Here either $A=0$ or $B=0$. \ For $A=0$
(outgoing case):%
\begin{eqnarray*}
\partial _{T}I+\partial _{X}I &=&0 \\
\frac{1}{4M}(X\partial _{T}I+T\partial _{X}I) &=&-E
\end{eqnarray*}%
The first equation implies the general solution of $I=h(X-T)$ and the second
in turn leads to

\begin{eqnarray*}
4ME &=&(X-T)h^{\prime }(X-T) \\
h^{\prime }(X-T) &=&\frac{4ME}{(X-T)}
\end{eqnarray*}%
which has a simple pole at the black hole horizon $X=T$. Setting $\eta =X-T$
we have

\begin{equation}
h^{\prime }(\eta )=\frac{4ME}{\eta }  \label{hprime}
\end{equation}%
Integrating (\ref{hprime}) around the pole at the horizon (doing a half
circle contour) implies%
\begin{equation*}
\func{Im}I_{out}=4\pi ME
\end{equation*}%
for outgoing particles.

For the incoming case $B=0$ and so

\begin{eqnarray*}
\partial _{T}I-\partial _{X}I &=&0 \\
\frac{1}{4M}(X\partial _{T}I+T\partial _{X}I) &=&-E
\end{eqnarray*}%
The first equation implies the general solution $I=k(X+T)$ and so the second
leads to

\begin{eqnarray*}
-4ME &=&(X+T)k^{\prime }(X+T) \\
k^{\prime }(X+T) &=&\frac{-4ME}{(X+T)}
\end{eqnarray*}%
Note that this equation does not have a pole at the black hole horizon $X=T$%
. \ Hence for incoming particles

\begin{equation*}
\func{Im}I_{in}=0
\end{equation*}%
and so Prob$[in]=1$ like in the Painlev\'{e} case. The final result for the
tunnelling probability is

\begin{equation*}
\Gamma =\frac{\text{Prob}[out]}{\text{Prob}[in]}=\exp [-2\func{Im}%
I_{out}]=\exp [-8\pi ME]
\end{equation*}%
and we see that the Hawking Temperature $T_{H}=\frac{1}{8\pi M}$ is
recovered in the massless case.

In the massive case we must use equations (\ref{defenergy}), (\ref{kruskal1}%
) and (\ref{kruskal2}) to solve for $\frac{A}{B}$. \ A straightforward
calculation yields%
\begin{equation}
\frac{A}{B}=\frac{-4ME\pm \sqrt{16M^{2}E^{2}+m^{2}f(r)(X^{2}-T^{2})}}{\sqrt{%
f(r)}m(X+T)}  \label{ksab}
\end{equation}%
where we note as the black hole horizon ($X=T$) is approached that either $%
\frac{A}{B}\rightarrow 0$ or $\frac{A}{B}\rightarrow \frac{-4ME}{\sqrt{f(2M)}%
mT}=\frac{-4ME}{\sqrt{f(2M)}mX}$. \ Subtracting (\ref{kruskal1})$/A$ from (%
\ref{kruskal2})$/B$ leads to 
\begin{equation*}
\partial _{T}I=-\partial _{X}I\frac{(1-(\frac{A}{B})^{2})}{(1+(\frac{A}{B}%
)^{2})}
\end{equation*}%
and so from (\ref{defenergy}) we obtain

\begin{equation}
\partial _{X}I=\frac{4ME(1+(\frac{A}{B})^{2})}{\left[ X(1-(\frac{A}{B}%
)^{2})+T(1+(\frac{A}{B})^{2})\right] }  \label{generalsol}
\end{equation}%
where $\frac{A}{B}\rightarrow 0$ at $X=T$.

From eq (\ref{ksab}) we find that

\begin{equation*}
\lim_{X\rightarrow T}\left[ X(1-(\frac{A}{B})^{2})+T(1+(\frac{A}{B})^{2})%
\right] =0
\end{equation*}%
and 
\begin{equation*}
\lim_{X\rightarrow T}\frac{\partial }{\partial X}\left[ X(1-(\frac{A}{B}%
)^{2})+T(1+(\frac{A}{B})^{2})\right] =\lim_{X\rightarrow T}\left[ (1-(\frac{A%
}{B})^{2})+2(X+T)(\frac{A}{B})\frac{\partial }{\partial X}(\frac{A}{B})%
\right] =1
\end{equation*}%
Consequently $\partial _{X}I$ has a simple pole at the black hole horizon
implying $\func{Im}I_{out}=4\pi ME$ in the massive case. \ Not that when $%
\frac{A}{B}\rightarrow \frac{-4ME}{\sqrt{f(2M)}mT}$ then $\partial _{X}I$
does not have a pole at the horizon, implying that $\func{Im}I_{in}=0$. \
The rest of the calculation proceeds as before, and we recover the Hawking
temperature in the massive case.

\subsection{Conclusions}

We have shown for the first time that computing the Unruh and Hawking
temperatures using the tunnelling method holds for fermions. \ Comparatively
few demonstrations that fermionic radiation has the same temperature as
scalar radiation due to the presence of these horizons appears in the
literature \cite{2d bh fermion}-\cite{Kinnersley BH}\textbf{.} \ These all
involve either lower dimensional calculations of the Bogoliubov
transformation \cite{2d bh fermion} or use of the GTCT \cite{gtct}-\cite%
{Kinnersley BH} to calculate fermion radiation from evaporating black holes.
\ We have shown that computation of black hole temperature for fermion
emission using tunnelling methods is relatively simple and straightforward
to compute.

For accelerated observers using Rindler coordinates we found the expected
Unruh temperature. \ We also applied fermion tunnelling to a general static
spherically symmetric black hole metric in both Schwarzschild and Painlev%
\'{e} form, and found that the usual Hawking temperature is recovered. \
That this situation does not depend on coordinate singularities was
demonstrated by showing the same results hold for the Kruskal-Szekers
metric. Our results indicate not only that the tunnelling method is robust,
but that it can indeed be understood as a physical phenomenon.\textbf{\ }

Extending fermion tunnelling to rotating spacetimes in which the emitted
particles have orbital angular momentum would be a natural next step.
Computing fermion tunnelling in the background of the Kinnersley metric is a
natural step. Based on the emission probability results from the Kinnersley
Black hole \cite{Kinnersley BH}, we expect that the final tunnelling
probability should be of the form $\exp (-\frac{1}{T_{H}}(E-\Omega
_{H}J_{\phi }+C))$, where $C$\ parametrizes the coupling between the spin of
the field and the angular momentum of the black hole. \ Extending fermion
tunnelling to dynamical black holes such as Vaidya or those used in \cite%
{dynamicalbh} would be a logical next step.\textbf{\ }Computing corrections
to the tunnelling probability by fully taking into account conservation of
energy will yield corrections to the fermion emission temperature. \ In
various scalar field cases this is inherent in the Parikh/Wilczek tunnelling
method \cite{Parikh}, \cite{Vagenas2}-\cite{charged} and can be incorporated
into the Hamilton-Jacobi tunnelling approach \cite{Vagenas1}. \ Another
avenue of research is to perform tunnelling calculations to higher order in
WKB (in both the scalar field and fermionic cases) in order to calculate
grey body effects. The possibility of calculating a density matrix for the
emitted particles via the tunnelling approach in order to calculate
correlations between particles is another interesting line of research. Work
on these areas is in progress.

{\Huge \bigskip }

{\Huge \medskip Acknowledgements}

This work was supported in part by the Natural Sciences and Engineering
Research Council of Canada.{\Huge \bigskip }

\end{document}